# Induced transparency in EXAFS as cuprate superconductors go through $T_c$

J. G. Chigvinadze[(1,2)], G. I. Mamniashvilli[(1,2)], J.V. Acrivos[(2,3)]

[(1)]E. Andronikashvili Institute of Physics, 0177, Georgia, [(2)]SJSU, CA 95192-0101

**ABSTRACT**

The increased transmission, observed in the EXAFS region of their X-ray absorption spectra, as cuprate materials go through the superconducting transition temperature $T_c$ is correlated with an increase in Abrikosov Vortex expulsion in zero magnetic field as the temperature T approaches $T_c$.

## INTRODUCTION

It has been known for some time that the formation of quantized vortices in rotating superfluid $^4$He induces transparency in its absorption spectrum [1]. The purpose of this work is to determine if the same effect can be observed in cuprate superconductors due to the formation of superconducting Abrikosov vortices [2]. Magnetic flux expulsion from a bulk superconductor is related to the magnitude of the external magnetic field. When it goes through zero the disintegration of Abrikosov vortex lattice is a cooperative dynamic phenomenon.

The rf power dissipation, as the magnetic field applied to oxide superconductors goes through zero [3-7] is a measure of flux expulsion [2]. X-ray absorption, XAS studies of superconducting cuprates at the Cu:K-edge, Ba:$L_{3,2}$-edges, Nd:$L_3$-edge and O:K-edge [8-11] have observed increased transmission as the sample passes through $T_c$ in the extended X-ray absorption spectra, EXAFS region. This work correlates the increased flux expulsion observed in going through $T_c$ and the decrease in the kinetic energy absorption in the EXAFS region of XAS.

## EXPERIMENTAL

The flux expulsion from a cuprate superconductor as the external magnetic field goes through zero is detected by the induced emf in an rf coil versus temperature T, as described previously [3, 4]. The detector is a Varian Wide Line nmr spectrometer (Fig. 1a). The sample preparation (Fig. 1b), analysis of the recorded data (Fig. 1c), and field calibration by low field electron spin resonance, esr of a free radical with line width less than an oe has been described previously [3-5]. The induced emf, $A_Y$ is recorded versus time t as the static field, $B_z$ is cycled through zero in a given period of time t (Fig. 1, 2). The detector ac field amplitudes (Fig. 1a) are $B_1$, $B_{zm}$ ~ 0.05 œ ~ $H_{c1}$. The powder cuprates $YBa_2Cu_3O_7$ (YBCO) and its derivatives $Nd(Ba_{0.95}Nd_{0.05})_2Cu_3O_7$ (Nd1.1-YBCO) and $(Y_{0.2}Ca_{0.8})Sr_2Cu_2(Tl_{0.5}Pb_{0.5})O_7$, (RSL-0.2) were ground, selected to a homogeneous particle size between 4 and 5 μm by sieves, and dispersed into an equal volume ratio of mixed 5 minute epoxy and filled into 2 inch long, 1mm id Pyrex tubes open on both sides, in less than a minute. They were then placed to cure in an external orienting field $H_o$ = 9 T in different geometries (Fig. 1b) with respect to the sample tube axis y' for at least 30 minutes. An SEM of the YBCO in epoxy matrix confirms the 50:50 distribution, but shows that smaller particles were broken off in the epoxy matrix (Figure 1b). The measurements are carried out by placing the epoxy matrix cylinder in a Dewar with the y' axis parallel to Y of the nmr probe (Fig. 1a, b). Liquid $N_2$ is poured into the Dewar, and the temperature T is recorded simultaneously with the probe emf output, using a copper-constantan thermocouple in contact with the sample (Fig. 1a, b and 2a, b). $T_c$ = 92.2 K for the YBCO in epoxy matrix was determined from the copper-constantan thermocouple calibration at the temperature where the change in temperature, T versus time t, dT/dt = 0, i.e., when the sample bulk heat capacity diverges (Figure 2a). The Varian spectrometer was operated at $\nu$ = 8 MHz with a modulation frequency $\nu_m$ = 400 Hz. This can detect an absorption signal derivative when the magnitude of the sweep field satisfies the resonance condition, but none was detected for $T > T_c$ (Fig. 2a). However, there is an increase in the flux expulsion signal, observed at 77 K by a factor of two when the sweep field increases from $B_z$ = ± 5oe to $B_z$ = ± 10oe and ± 50oe.

XAS spectra at the Cu:K-edge, Ba:$L_{3,2}$-edges, Nd:$L_3$-edge, O:K-edge of the different cuprate superconductors were measured at the Stanford Synchrotron Radiation laboratory (SSRL) and the Lawrence Berkeley Advanced Light Source (LBL ALS) as the sample goes through $T_c$ (Fig. 3) [7-9]. The Absorbance A of samples contained in a He Dewar are detected as T increases/decreases through $T_c$ at different rates. Increased transmission is observed in the EXAFS region of XAS as the superconductor goes through $T_c$ (e.g., Fig. 3a, b). The structural changes derived from the EXAFS analysis [8-11] indicate that within the error of the measurements/analysis the bond distances in cuprates do not change through $T_c$, as expected for a 2$^{nd}$ order phase transition. The XAS measurements are taken in the earth's magnetic field.

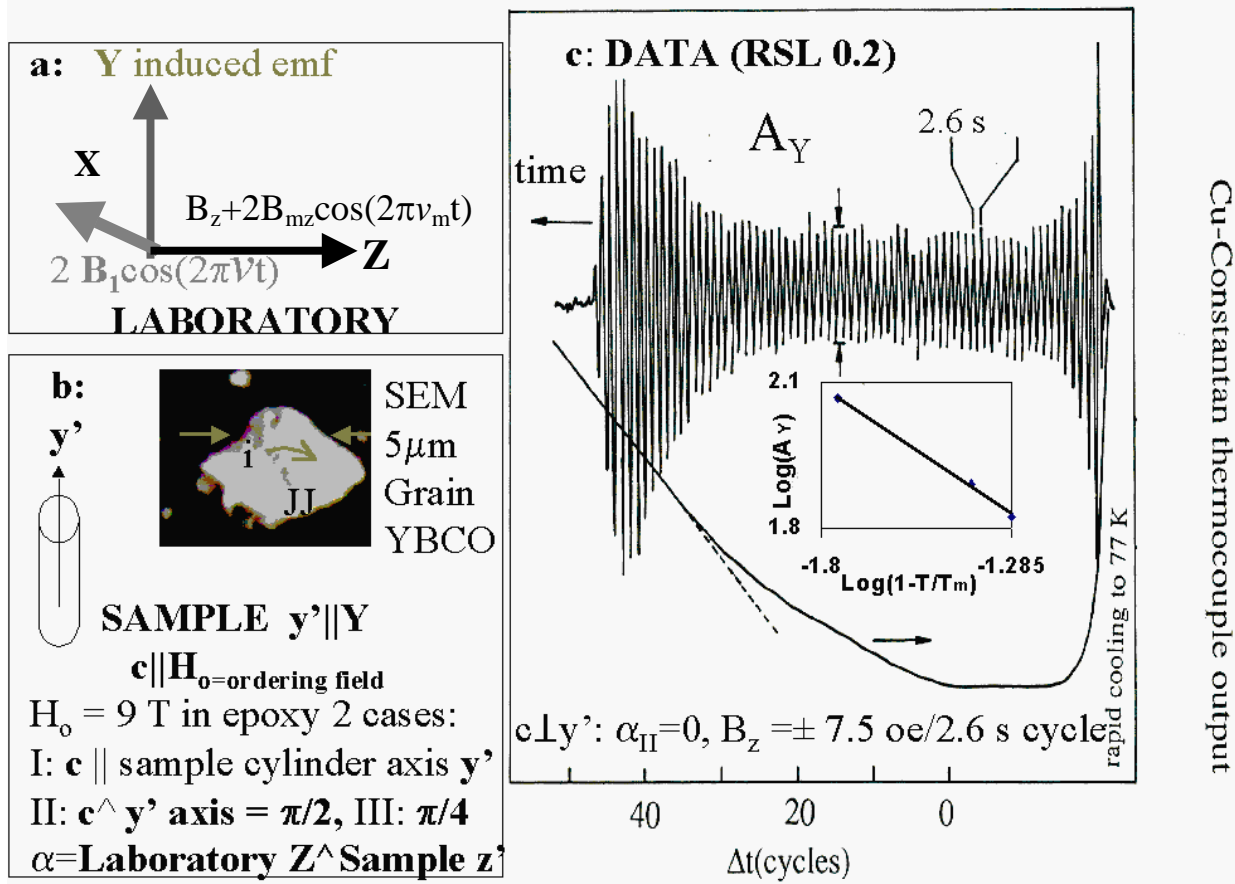

**FIG. 1:** *Preparation of flux lattice and EMF measurement of flux expulsion:* **(a)** *Wide line Varian nmr spectrometer parameters: $v = 8$ MHz and $v_m = 400$ MHz in the geometry shown.* **(b)** *Three cases of sample orientation is given by the angle the cylinder axis y' makes with the orienting field $H_o$. Cases I, II, or III identify $y'\wedge H_o = 0, \pi/2$ and $\pi/4$. SEM of the YBCO matrix shows a sieve selected particle plus smaller ones produced during the orientation in epoxy.* **(c)** *$A_Y$ is recorded simultaneously with the output of a copper-constantan thermocouple in mV, and calibration gives an absolute temperature to better than $T \pm 2$ K. The insert shows data for a well characterized cuprate superconductor with $T_c \approx 105$ K [5b]. $A_Y$ diverges as $T \to T_m$ with a power law $|1-T/T_m|^{-(0.5\pm 0.1)}$. The calibration is in agreement with the reported $T_c = 105$ K [ref. 5b].*

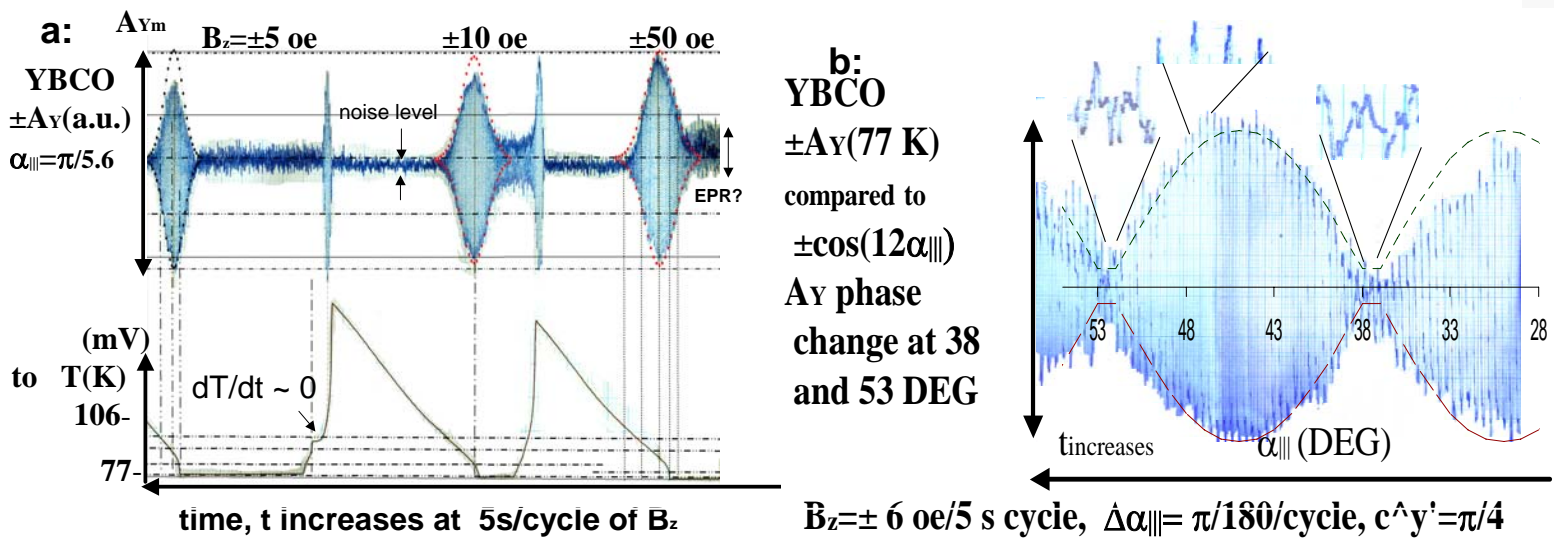

**FIG. 2:** *$A_Y$ versus temperature T for the YBCO in epoxy matrix. The dotted curves are fits to a gaussian distribution of particles for the enhanced flux expulsion near $T_c$:* **(a)** *Effect of the magnetic field sweep width $B_z = \pm 5, 10, 50$ oe on $A_Y(\alpha_{|||} = \pi/6)$.* **(b)** *$A_Y(\alpha_{|||})$ taken at 1 DEG angle increments per field cycle, at 77 K over a quadrant for the YBCO matrix showing a 12 fold symmetry that goes out of phase above 60 DEG. The change in phase as $A_Y$ goes through zero (expanded at 48 and 53 DEG) indicates that the induced emf changes sign as it goes through zero. Table gives the half widths at half height, HWHH.* **(c), (d)** *Angular dependence of $A_{Y,max}(\alpha_i)$ for cases i = I and II (Fig. 1b). Comparison of the data points with $\cos(n\alpha)$ and $\cos^2(n\alpha)$ versus $\alpha_i$, for n-fold rotation symmetry indicates a better fit to the former.*

**c:** 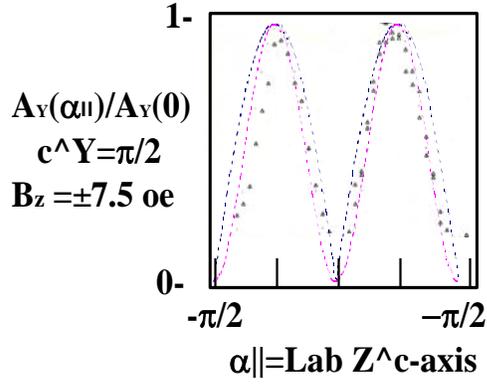

$A_Y(\alpha_{||})/A_Y(0)$
$c\wedge Y=\pi/2$
$B_z =\pm 7.5$ oe

$\alpha_{||}$=Lab Z$\wedge$c-axis

**d:** 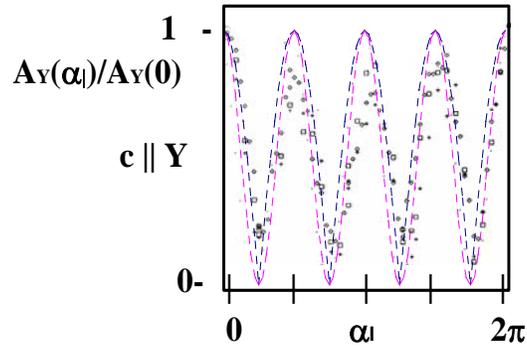

$A_Y(\alpha_|)/A_Y(0)$
c || Y

$\alpha_|$

**Table I:** Line widths (HWHH) of critical oscillation as $\pm B_z$ goes through zero above $T_c$ (FIG.1c, 2a).

| Sample | Case | $\alpha$ | $B_z$ (oe) | $T_{HWHH}-T_m$ (K) |
|---|---|---|---|---|
| RSL:0.02 | II | 0 | 7.5 | 2.5 |
| YBCO | III | $\pi/5.6$ | 5 | 8.6 |
| YBCO | III | $\pi/5.6$ | 10 | 10.6 |
| YBCO | III | $\pi/5.6$ | 50 | 10.6 |

**a:** Nd(Ba$_{0.95}$Nd$_{0.05}$)$_2$Cu$_3$O$_7$: 2μm powder in BN
Nd: $(1s)^2(2p_{3/2})^2$.. $\Leftrightarrow$ Nd: $(1s)^2(2p_{3/2})$..nd

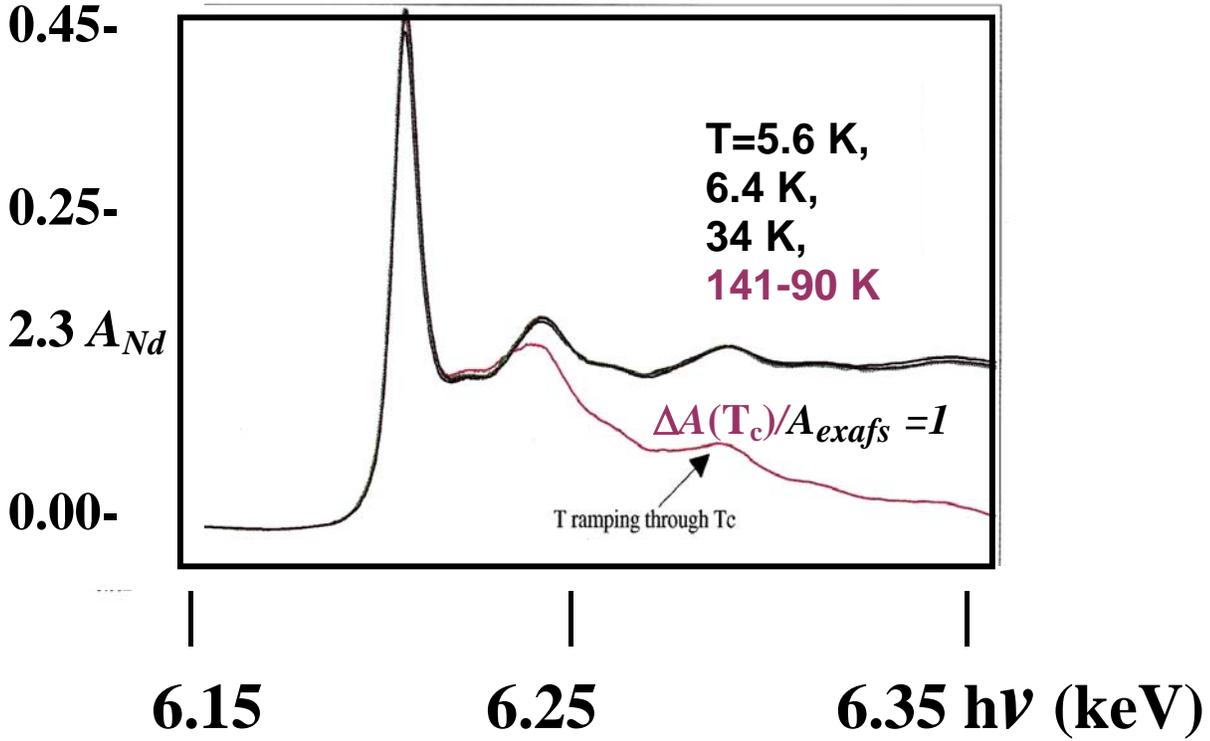

T=5.6 K, 6.4 K, 34 K, 141-90 K

2.3 $A_{Nd}$

$\Delta A(T_c)/A_{exafs} =1$

T ramping through Tc

6.15    6.25    6.35 h$\nu$ (keV)

**FIG. 3:** *Increased transmission in XAS of cuprate superconductors as the temperature increases/decreases through $T_c$ [8-11]: (**a**) Absorbance of Nd(Ba$_{0.95}$Nd$_{0.05}$)$_2$Cu$_3$O$_7$, 2 μm powder dispersed in BN near Nd$_{L3\text{-edge}}$. (**b**) Absorbance of YBa$_2$Cu$_3$O$_7$ single crystal (of thickness $d_c \sim 44$ μm) near the Cu$_K$ edge. The single crystal shows two transitions to superconductivity near 92 and 60 K.*

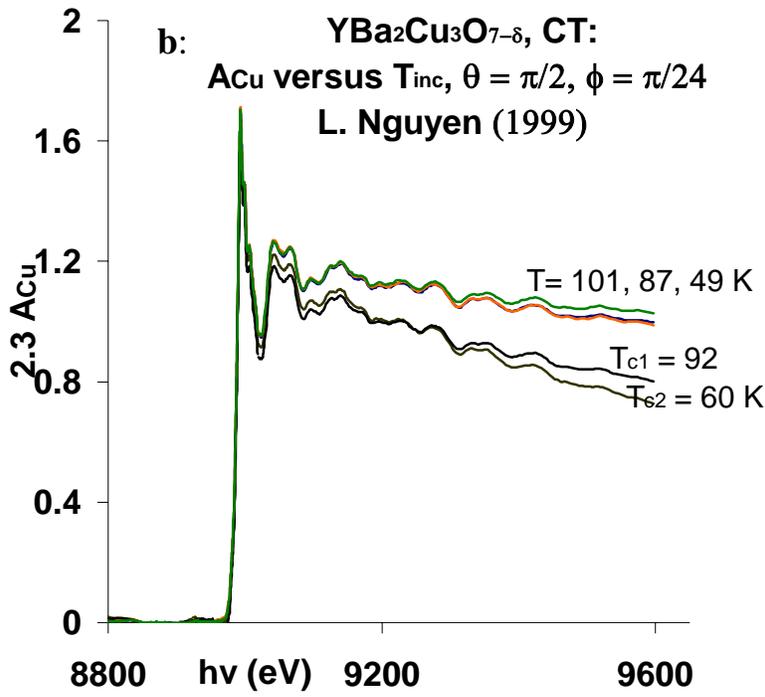

b: $YBa_2Cu_3O_{7-\delta}$, CT:
$A_{Cu}$ versus $T_{inc}$, $\theta = \pi/2$, $\phi = \pi/24$
L. Nguyen (1999)

T= 101, 87, 49 K
$T_{c1}$ = 92
$T_{s2}$ = 60 K

## DISCUSSION

Theory indicates that when the applied external magnetic field goes through zero the Abrikosov vortex lattice disintegrates and vortices start to escape the bulk superconductor [2]. The escape of magnetic flux may then detected by the induced emf signal (Fig. 1c, 2a, b).

As T increases/decreases through $T_c$, the induced emf amplitude $A_Y$ goes through a maximum $A_{Y,max}$ at a temperature $T_m$ that is identified by the maximum flux expulsion signal (Fig. 1, 2). $T_m$ is above the value of $T_c$ reported by other measurements [5b] and the value recorded in the same experiment on cooling/warming past $T_c$, determined from the change in T versus time, dT/dt as the bulk superconductor heat capacity diverges [Fig. 2a]. In samples with a sharp transition to superconductivity $A_Y$ diverges as $|1-T/T_m|^{-(0.5 \pm 0.1)}$ (insert Fig. 1c).

The amplitude of the enhanced flux expulsion signal versus temperature for samples with a large particle size distribution (Fig. 1b) show a gaussian shape (Fig. 2a). The RSL 0.2 in epoxy matrix has a very uniform size distribution [3b] with a HWHH of 2 K whereas the YBCO in epoxy matrix with different size particles (Fig. 1b) is fitted to a gaussian shape with a HWHH of 8 to 10 K (Fig. 2a), suggesting that the particle size is associated with the critical flux expulsion.

There is no indication of a magnetic resonance signal at temperatures above where the critical oscillations vanish for all the sweep fields. However an increase in the flux expulsion amplitude at 77 K, by a factor of two when the sweep field increases to $B_z = \pm 10$ and 50 oe from $\pm 5$ oe suggests that there is a flux expulsion enhanced resonance absorption satisfied for 10 oe > $B_z$ > 5 oe. This gives the limit for the Landé factor g < 1.1 suggesting an orbital angular momentum g = 1 or S = 0, resonance absorption. The resonance absorption is of opposite sign as that for the flux expulsion signal, and produces a kink in the signal (Figure 2a, $B_z = \pm 6$ oe). The width of the fitted gaussian shape for $A_{Y,max}$ versus temperature is 2 K broader for field sweeps $B_z = \pm 50, 10$ oe than for $B_z = \pm 5$ oe (Table 1), indicating that the gaussian shape is also determined by processes associated with the external sweep field. The linear AC response of the flux lines created by external magnetic fields normal to the layers of the layered superconductor has been calculated [16]. Two-dimensional (2D) vortices were related to the Berezinskii-Kosterlitz-Thouless [13] transition in high-$T_c$ superconductors and thermal fluctuations in the shifts of 2D-vortices forming flux lines, and the flow is subject to the Fokker-Planck relations, including pinning in small magnetic fields. This may be the cause of the line widths dependence on the field sweep (Fig. 2a). Other processes may be induced by the action of two radiation fields used. A. N. Lykov [14] has observed the emission of coherent microwave radiation from superconducting Nb films in the frequency range up to 600 MHz by the superposition of two alternating magnetic fields directed perpendicular to film surface. The slow varying field sets up a vortex structure in the film, while a second high-frequency electromagnetic field provides a synchronization of Abrikosov vortex motion. The simultaneous action of the fields results in either the amplification or generation of electromagnetic radiation. Harmonic mixing of the radiation was also detected. The question remains whether the critical oscillations are related to the macroscopic mixed state (MACMS), corresponding to a mixture of macroscopic large supeconducting and normal domains, that may obtain in magnetic field $B_z$. Moscchalkov [17] has found that in the H-T phase diagram MACMS lies between high temperature normal state and low temperature Abrikosov vortex state. The irreversibility line corresponds in this case to the topological cross-over from MACMS with isolated singly-connected superconducting "droplets", surrounded by the normal phase, to be sponge-like MACMS with multiply-connected superconducting "body", surrounding isolated normal-state domains, which finally decay into Abrikosov vortices, as temperature goes down. The fact that enhanced critical expulsion of Abrikosov vortices is observed above/below $T_c$, in low dimensional cuprates must be taken into account in say a Berezinskii-Kosterlitz-Thouless approach [13] to explain the many new phenomena in these materials above the nominal transition temperature to superconductivity (Fig. 1, 2a) and others [14-16].

The twofold symmetry of $A_{Y,m}(\alpha_{\parallel})/A_{Y,max}$ versus the angle orientation in $B_z$, $\alpha_{\parallel} = c \wedge B_z$ is reasonable, and may be interpreted as a dependence of the magnitude of magnetic field penetration in the direction of the c-axis, i.e, $B_c = B_z \cos(\alpha_{\parallel})$.

The fourfold symmetry shown by $A_{Y,m}(\alpha_{\parallel,c\parallel Y})/A_{Y,max}$ (Fig. 2d) versus orientation $\alpha_{\parallel}$ = (ab-plane axis)$\wedge B_z$ is surprising but may be understood by the ease of YBCO in-plane flux penetration if the in-plane YBCO conductivity is of $d_{xy}$ symmetry for a dominating particle in the matrix as determined by photoemission [12].

The 12-fold symmetry of $A_{Y,m}(\alpha_{\parallel\parallel,c\wedge Bz}=\pi/4)/A_{Y,max}$ versus $\alpha_{\parallel\parallel} = (\pi/4$ axis to c$)\wedge B_z$ (Fig. 2a, b) though confirmed by repeated measurements is hard to reconcile, but it may shine light on the phenomenon. It may result as a combination of the fourfold in-plane $d_{xy}$ conduction symmetry times a three fold symmetry for the formation of Abrikosov vortices [2] affecting the inter-plane interactions. A six-fold symmetry in the inter-plane coupling and the 2D-3D re-ordering transition of vortex lattices, observed in a disordered anisotropic superconductor, $Ba_2Sr_2CaCu_2O_8$, (BSSCO) was explained by both long and short range pancake-like interactions between the planes [12]. This effect should be enhanced when **c^B$_z$** ≠ 0 in cases II and III above.

The excess energy in XAS (Fig. 3) h$\nu$ (h = Planck constant and $\nu$ = frequency) greater than a core excitation energy h$\nu_0$, is converted by the photoelectron to kinetic energy,

$$p^2/2\, m_e = h(\nu-\nu_0) \qquad (1)$$

where **p** = **k** h/2$\pi$ is the momentum and $m_e$ the photoelectron mass. Careful XAS measurements versus temperature, aimed to measure the effects produced on the absorbance, $A$ near the phase transition of superconducting cuprates indicate an increased transmission only in passing through $T_c$, in the EXAFS region [7-11]. There is very little difference between the EXAFS spectra at temperatures above or below $T_c$ (Fig. 3). The flux expulsion due to the formation of vortices observed at T ~ 77 K (e.g., Fig. 1c, 2a) does not appear to affect the XAS. However, if photoelectrons become involved in Abrikosov vortices, the matrix element for EXAFS absorption must be different in the superconducting state. A divergence in the number of flux vortices may enhance the effect that results in the increased transmission in the EXAFS region of XAS, and this may be observed only as the temperature goes through $T_c$ (Fig. 3). This decrease in absorbance suggests the presence of an effect on EXAFS transmission by Abrikosov vortices [2] similar to the transparency produced by vortices in $^4$He [1]. It is important to note that 100 % transparency in the EXAFS region is induced in 2$\mu$m particles (Fig. 3a) whereas only 20 % transparency is induced in a bulk single crystal (Fig. 3b).

## CONCLUSIONS

Magnetic flux expulsion from the bulk of a superconductor is related to the external magnetic field going through zero and, consequently, with the disintegration of Abrikosov vortex lattice which is a cooperative dynamic phenomenon that affects the EXAFS spectrum when the number diverges beyond a critical value near $T_c$, necessary to induce a transparency in the absorbance.

## ACKNOWLEDGEMEMNTS

J. Chigvinadze and G. Mamniashvili thank the International Scientific and Technology Center for the ISTC Grant G-389 for their research visit to San Jose' State University.